\documentclass{aastex62}
\usepackage{booktabs}
\usepackage{rotating}
\usepackage{natbib}
\usepackage{xspace}

\newcommand{\lt}{\ensuremath <}

\newcommand{\celerite}{\texttt{celerite}\xspace}
\newcommand{\bls}{\texttt{astropy.stats.BoxLeastSquares}\xspace}
\newcommand{\kepler}{\textit{Kepler}\xspace}
\newcommand{\ktwo}{\textit{K2}\xspace}
\newcommand{\tess}{\textit{TESS}\xspace}
\newcommand{\lk}{\texttt{lightkurve}\xspace}
\newcommand{\exoplanet}{\texttt{exoplanet}\xspace}
\newcommand{\starry}{\texttt{starry}\xspace}

\newcommand{\spitzer}{\textit{Spitzer}\xspace}
\newcommand{\tls}{\texttt{tls}\xspace}

\graphicspath{{./}{}}

\received{2019 April 23}
\accepted{2019 June 18}
\published{2019 July 18}
\submitjournal{ApJ}

\shorttitle{Four Small Planets Buried in \ktwo Systems}
\shortauthors{Hedges et al.}

\begin{document}

\title{Four Small Planets Buried in \ktwo Systems: What Can We Learn For \tess?}

\author{Christina Hedges}
\affil{Bay Area Environmental Research Institute, P.O. Box 25, Moffett Field, CA 94035, USA}
\email{christina.l.hedges@nasa.gov}

\author{Nicholas Saunders}
\affil{Bay Area Environmental Research Institute, P.O. Box 25, Moffett Field, CA 94035, USA}

\author{Geert Barentsen}
\affil{Bay Area Environmental Research Institute, P.O. Box 25, Moffett Field, CA 94035, USA}

\author{Jeffrey L. Coughlin}
\affil{SETI Institute, 189 Bernardo Ave, Suite 200, Mountain View, CA 94043, USA}

\author{Jos\`e Vin\'icius de Miranda Cardoso}
\affil{Federal University of Campina Grande, Department of Electrical Engineering, Brazil}

\author{Veselin B. Kostov}
\affil{NASA Goddard Space Flight Center, Greenbelt, MD 20771, USA}
\affil{SETI Institute, 189 Bernardo Ave, Suite 200, Mountain View, CA 94043, USA}

\author{Jessie Dotson}
\affil{NASA Ames Research Center, Moffett Field, CA 94035, USA}

\author{Ann Marie Cody}
\affil{Bay Area Environmental Research Institute, P.O. Box 25, Moffett Field, CA 94035, USA}

\begin{abstract}

The \kepler, \ktwo, and \tess~ missions have provided a wealth of confirmed exoplanets, benefiting from a huge effort from the planet hunting and follow-up community. With careful systematics mitigation, these missions provide precise photometric time series, which enable detection of transiting exoplanet signals. However, exoplanet hunting can be confounded by several factors, including instrumental noise, search biases, and host star variability. In this letter, we discuss strategies to overcome these challenges using newly-emerging techniques and tools. We demonstrate the power of new, fast open source community tools (e.g., \lk, \starry, \celerite, \exoplanet), and discuss four high signal-to-noise exoplanets that showcase specific challenges present in planet detection: \textit{K2-43c}, \textit{K2-168c}, \textit{K2-198c}, and \textit{K2-198d}. These planets have been undetected in several large \ktwo planet searches, despite having transit signals with SNR$\geq10$. Two of the planets discussed here are new discoveries. In this work we confirm all four as true planets. Alongside these planet systems, we discuss three key challenges in finding small transiting exoplanets. The aim of this paper is to help new researchers understand where planet detection efficiency gains can be made, and to encourage the continued use of \ktwo archive data. The considerations presented in this letter are equally applicable to \kepler, \ktwo, and \tess, and the tools discussed here are available for the community to apply to improve exoplanet discovery and fitting.
\end{abstract}


\section{Introduction}
\label{sec:intro}

The \kepler mission \citep{borucki} has led to the detection of thousands of exoplanets, which have been used to further our understanding of planet occurrence rates and planet formation. After the loss of a second reaction wheel caused the \kepler spacecraft to lose fine pointing ability, the \kepler mission began a new phase, named \ktwo \citep{howell}. The \ktwo mission has since lead to the detection of more than 350 confirmed, transiting planets, continuing the work from the \kepler mission.

The new \tess mission \citep{ricker}, launched in 2018, is building on the legacy of \kepler to provide an all sky survey of nearby planet systems. In many ways, the data from \tess and \ktwo are similar; both have short observing campaigns, are subject to sub-pixel spacecraft motion, and observe a significant number of young and active stars, in contrast to the original \kepler mission which focused on main sequence FGK stars and benefited from extremely-precise pointing stability. By better understanding our planet hunting biases from \ktwo, we can improve our efficiency in finding planets with both \ktwo and \tess, and ultimately increase our planet detection efficiency.

In this work we undertake a planet search in existing planet systems from early \ktwo campaigns, in order to demonstrate the power of new, open source software to overcome key problems in planet hunting searches with \ktwo data. This simple search was not designed to be complete. Instead, it was designed to highlight cases where new tools offer significant benefits to the community, inspired by a recent study in which we concluded that hundreds more planets remain to be discovered in the \ktwo data set \citep{dotson}. This paper aims to encourage the continued use of \ktwo archive data by helping new and current researchers understand where improvements in transiting planet searches can continue to be made.

 In this paper we discuss three systems with confirmed transiting planets, where we have identified small planets that have remained unreported. These smaller planets have evaded detection for approximately 4 years, despite several pipelines identifying the larger planets in the systems \citep{Vanderburg2016, crossfield, mayo}, and despite the unreported additional planets showing a high signal-to-noise ratio (SNR$\,>\,9$). We present 4 additional planets in 3 systems in the following sections, and discuss in detail the factors we have identified that contributed to their being unreported. The SNR of small transiting planet signals can be improved by applying the analysis modifications presented in this paper, increasing the likelihood of detection. If these simple methods are implemented in large pipeline searches for planets, they will increase planet detection efficiency, particularly for the \ktwo and \tess misisons.

 The challenges discussed here are equally important for planets in \tess data. Likewise, the methods and tools we have used to identify and fit these planets in \ktwo data are equally applicable to \tess data.  If unaddressed, these challenges may cause valuable multi-planet systems in \tess to also remain undetected.

 In brief we discuss three key reasons that these particular planets have remained undiscovered:

\begin{itemize}
    \item \textbf{High-frequency pointing jitter}: The first three \ktwo Campaigns experienced increased levels of high-frequency spacecraft motion. This intra-cadence noise is challenging to remove and can obscure \textbf{small} planet candidates by reducing the transit SNR.
    \item \textbf{Resonant planets}: Planets can naturally occur at orbital resonances, which cause peaks in the periodograms used to search for transiting signals to overlap with harmonics. Clipping out harmonics of significant peaks, or clipping transits, can reduce sensitivity to these resonant planets.
    \item \textbf{Stellar variability}: High amplitude stellar variability (e.g., from star spots) must usually be removed before searching for planets. If this variability is removed with an inadequate model, the residuals of the stellar variability model fit can be larger than any transiting planet signals, causing small planets to be lost in the noise.
\end{itemize}

In the following sections we present four planets, two of which are new discoveries, and two of which were previously identified as candidates in \cite{pope}, though unreported in the NASA Exoplanet Archive. In this work we confirm all four signals as true planet detections at greater than 95\% confidence. In Section \ref{sec:community} we discuss the community tools and techniques we have used to search and fit transiting planet signals. In Section \ref{sec:method} we first discuss our planet search and fitting methods. Each system is discussed in detail in Section \ref{sec:results}, alongside a discussion of the factors that previously obscured these planets from being detected. We present full transit model fits for all planets in the systems (both previously identified, and discovered by this work) and vetting statistics for each planet. The scripts used to analyze each of the systems are available online\footnote{\url{https://github.com/christinahedges/threemultis}}.

\section{Background: Community tools in \kepler, \ktwo, and \tess}
\label{sec:community}

The \ktwo mission provided the community with observations taken in $\sim80$ day campaigns, pointed towards the ecliptic plane. The roll motion of the spacecraft caused target motion relative to the detector of at most $2$ pixels and typically $\lesssim1$ pixel. This roll motion generated characteristic noise in \ktwo observations, caused by the Point Spread Function (PSF) of the star moving over sub-pixel sensitivity variations. Several attempts have been made to correct this noise, typically using one of three key methods: the Self Flat-Fielding Technique (SFF), Pixel Level Decorrelation (PLD), and Gaussian Process Detrending. These methods have been used in several community pipelines: e.g., K2SFF \citep{vanderburg} and K2P2 \citep{lund} use SFF; EVEREST \citep{luger_a, luger_b} uses PLD; and K2SC \citep{aigrain} and K2PHOT \citep{petigura2018} use Gaussian Process Detrending. These pipelines typically achieve a correction of the roll motion within a factor of 2-4 of the original \kepler precision.

 Several new community tools have become available in the past year for working with \kepler data, producing light curves, removing instrument systematics and stellar variability, and discovering and fitting planet signals. These new tools and methods enable us to overcome the problems outlined in section \ref{sec:intro} and discussed in detail in section \ref{sec:results}, and find these planets that were previously buried in the noise. There are many new tools available to work with \kepler, \ktwo, and \tess data\footnote{A full list of community tools for working with \kepler, \ktwo, and \tess data can be found at \url{https://docs.lightkurve.org/about/other_software.html}}. The specific tools we have used in this letter are listed below.

\begin{itemize}
    \item \lk\footnote{\url{https://docs.lightkurve.org}} \citep{lightkurve}: A new open source Python package to work with \kepler, \ktwo, and \tess data. Notably, \lk enables users to extract photometry using custom aperture masks, and remove motion systematics using tunable implementations of the SFF and PLD techniques.
    \item \bls\footnote{\url{http://docs.astropy.org/en/latest/stats/bls.html}} \citep{astropy:2013,astropy:2018}: A new, fast implementation of the Box Least Squares \citep[BLS;][]{BLS1, BLS2} planet-finding algorithm in Python.
    \item \exoplanet\footnote{\url{https://exoplanet.dfm.io}} \citep{exoplanet}: A set of tools for fitting exoplanet transits and Gaussian Processes to long term trends using \texttt{pymc3}\footnote{\url{https://docs.pymc.io/}} based on the \starry\footnote{\url{https://rodluger.github.io/starry/}} and \celerite\footnote{\url{https://celerite.readthedocs.io}} packages \citep{pymc3,starry,celerite}.
\end{itemize}

\section{Method}
\label{sec:method}

\subsection{Planet Search}
\label{sec:searching}

In an effort to identify planets that have remained unreported in the NASA Exoplanet Archive we used the following approach. We selected only confirmed planets hosts from early campaigns 0 through 8, resulting in 164 stellar systems to search. These early campaigns have been archived at MAST for approximately 4 years. We obtained the \kepler Pipeline \citep{jenkins2010} products from MAST, and used the Pre Data Search Conditioning Simple Aperture Photometry (PDCSAP) light curves of the existing planet host identified in the NASA Exoplanet Archive. We used a combination of Gaussian process (GP) detrending to remove long term trends  \cite[for discussion of GPs and their application to time series photometry see][]{celerite}, and the Self Flat Fielding (SFF) technique \citep[see][]{vanderburg} to remove short-term K2 roll motion systematics using \lk. We chose SFF for its simple implementation and speed. We generated hundreds of light curves for each target with varied motion detrending parameters and best fit GP hyperparameters. We removed the signals of known planets by masking transits, and replaced them with Gaussian noise with the same standard deviation as the out-of-transit light curve. We then used \bls to identify targets where there were signals of an additional transiting planet in a significant number of our hundreds of light curves.

Candidate transiting planets were identified by stacking BLS periodograms for each of the hundred light curves, and searching for over-densities that indicated a transiting signal in at least 10\% of the detrended light curves, resulting in $\lesssim$ 10 candidate systems. These candidates were then inspected by eye, and systems with transit signals with SNR$\geq$4 were invesigated in detail. We identified three systems with significant power in the BLS periodogram, which demonstrated common problems that are encountered in planet searches. We subsequently confirmed each of these planets (see Section \ref{sec:results}). Once our candidates had been identified, we used a separate method to produce more accurate light curves, as discussed in Section \ref{sec:fitting}. 

Our search of \ktwo targets is by nature not complete, and is designed only to identify common pitfalls that have obscured planets from being identified in the  \ktwo dataset.  Our search is restricted only to confirmed planet systems from early campaigns. We vary only two of our detrending parameters, however it would be possible to vary more parameters to improve completeness, including detrending method, pixel aperture size and long-term detrending method. In this letter we do not aim for completeness, and instead discuss three interesting case studies of planets that have been unreported, what factors have led to their obscurity, and how to alleviate these problems.

\subsection{Planet Fitting}
\label{sec:fitting}


Once we had identified systems with significant evidence of additional planets, with a permissive threshold of SNR$\geq4$, we recreated light curves using a slower, but more accurate approach. We use \exoplanet to simultaneously detrend spacecraft motion noise, using 2nd order PLD, and a Gaussian Process to remove stellar variability using a Mat\'ern 3/2 kernel. We remove any transits from this step, to ensure that transits do not inform our estimate of the stellar variability. We then use our GP to predict the stellar variability during transits, and an MCMC to estimate the uncertainties during transit, (which are slightly larger than the out-of-transit uncertainties). The uncertainties on each data point in the light curve have been marginalized over the uncertainities in our GP hyperparameters, meaning that they robustly capture our uncertainties from detrending. Preserving our uncertainties in our light curve allows us to accurately capture the uncertainties in our planet parameters.

Once we have our final light curve, we fit a multi-planet transit model using \exoplanet, and use an MCMC to estimate the uncertainties on our planet parameters. The final results of our transit fit are given in Table \ref{tab:planetparams}, with 1$\sigma$ uncertainties. For each transit fit we use literature values for stellar parameters, from \cite{dressing} and \cite{mayo} (see Table \ref{tab:planetparams}).


\section{Results}
\label{sec:results}

\subsection{K2-43}
\label{sec:k2-43}


\textit{K2-43} (EPIC 201205469) was observed in long cadence in Campaign 1 of the \ktwo mission (program GO1059; PI Stello), and discovered to host a large, short period planet in \cite{Vanderburg2016, crossfield}. In early \ktwo campaigns (C0-2), the spacecraft used a lower pointing control frequency which caused increased levels of intra-cadence motion compared to later K2 Campaigns \citep{k2handbook}. This motion caused a significant change to the apparent measured PSF shape from frame to frame, as this motion was coadded during a single candence. This apparent change in shape causes some motion detrending methods to fail. Methods such as SFF are fast and simple, and often provide excellent results. However in the presence of intra-cadence motion, SFF can fail to produce the most accurate light curve, as the slight change in PSF shape causes changes to the target centroid, which is crucial to the success of the SFF method. The PLD method performs better on the extremely short duration motion noise that is present in these early campaigns.

\begin{sidewaysfigure}[h]
    \centering
    \includegraphics[height=0.45\textwidth]{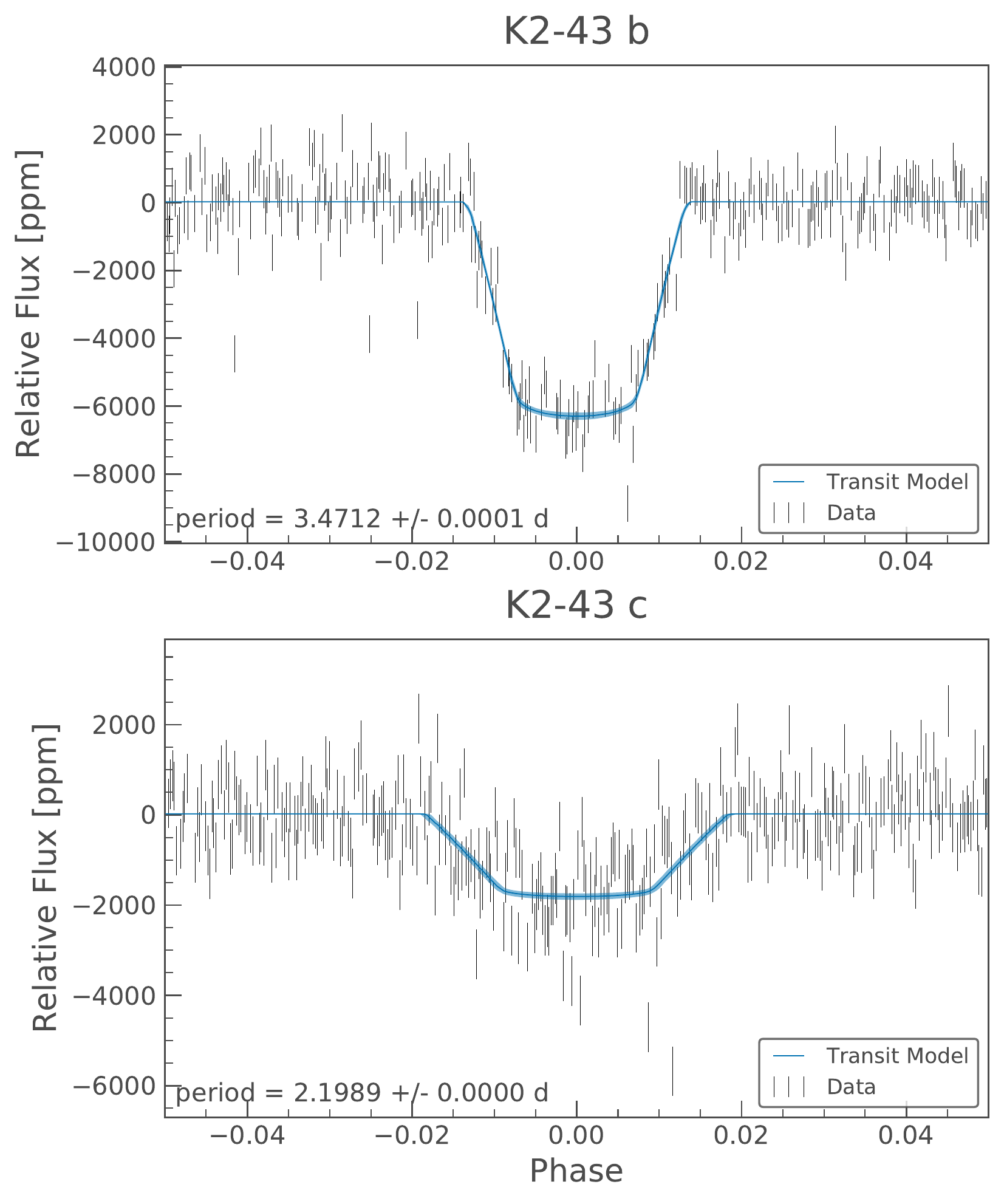}
    \includegraphics[height=0.45\textwidth]{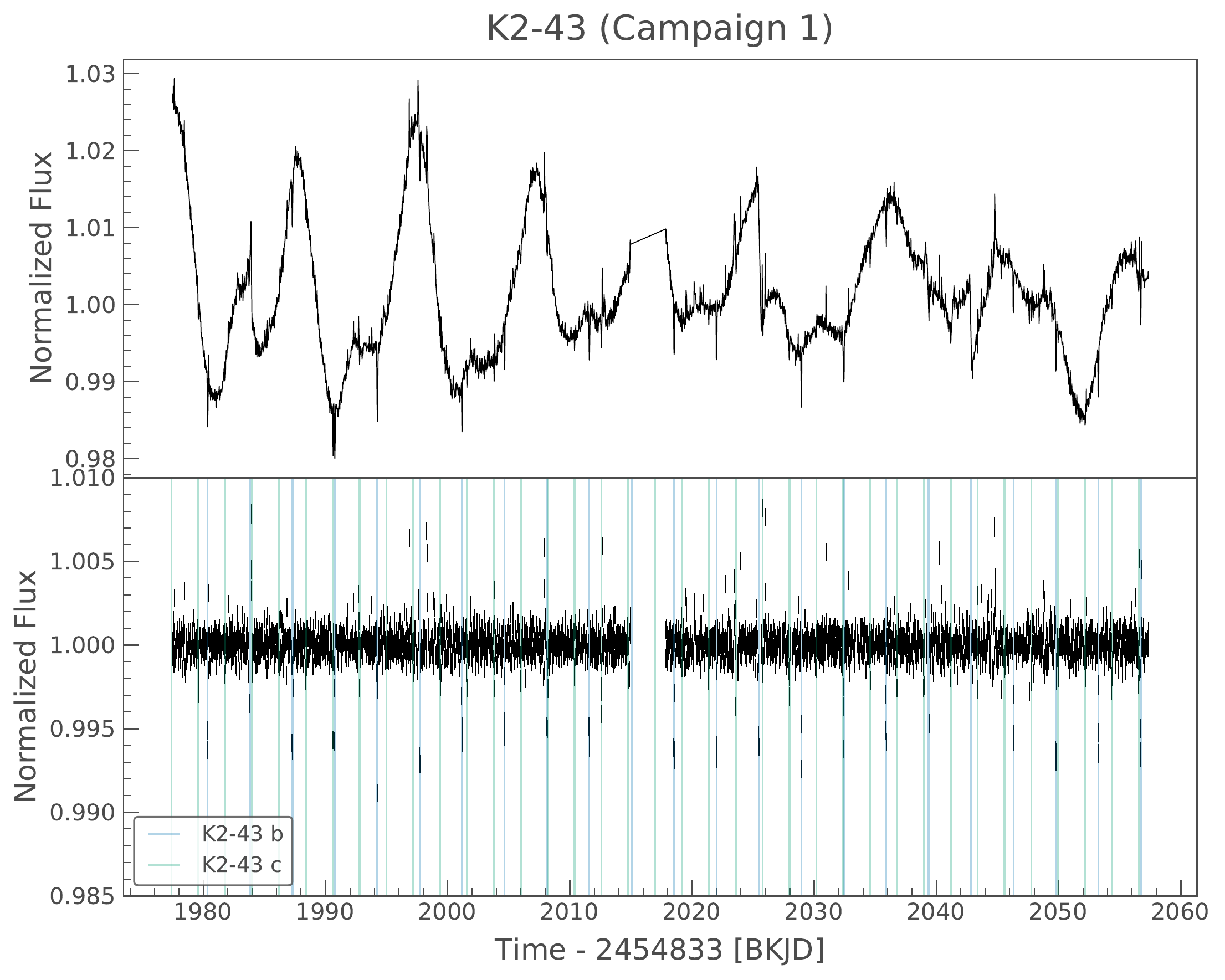}
    \caption{\textit{Left}: Folded transits of \textit{K2-43b} and \textit{c}, having simultaenously removed \ktwo motion systematics and long term stellar variability (see Section \ref{sec:fitting}. Our best fit planet model is show in blue alongside the 1$\sigma$ uncertainty. \textit{Right}: Light curves of \textit{K2-43}. \textit{Top}: Light curve with motion systematics corrected. Strong stellar varaibility due to spots is clearly evident. \textit{Bottom}: Light curve with both motion systematics and stellar variability removed. Transits of \textit{K2-43b} and \textit{c} have been highlighted.}
    \label{fig:k2-43}
\end{sidewaysfigure}

Early in the K2 mission, systematics-corrected light curves were provided to the community by the K2SFF Pipeline \citep{vanderburg} which uses the SFF technique. \cite{Vanderburg2016}, \cite{dressing}, and \cite{mayo} use light curves from or using the same process as \cite{vanderburg}, and \cite{crossfield} build their own SFF light curves using a similar approach but modeling correlations between spacecraft roll and flux with a different method. As such, each of these works used methods that were unable to optimally correct the intra-cadence motion. \textit{K2-43b} was identified in the SFF lightcurves, where the SNR of the transits of a second planet \emph{K2-43c} is low (SNR$\lt$9). The small transiting planet signal was not identified, as it was buried in the high-frequency motion noise experienced by early K2 campaigns. Using our method (described in Section \ref{sec:searching}) we identified weak evidence of a transiting planet \textit{K2-43c}. After identifying the low SNR signal, we built our refined light curve using the method discussed in section \ref{sec:fitting}, which greatly improved the SNR. By applying PLD detrending, the transit is identifiable as a SNR$>10$ signal. Upon reviewing the EVEREST light curves, which also use PLD, and we find that \textit{K2-43c} is also identifiable in the EVEREST community pipeline, which was released after K2SFF in 2016. igure \ref{fig:k2-43} shows the folded transits of \textit{K2-43b} and the new planet identified in this work, \textit{K2-43c} using a PLD detrending (see Section \ref{sec:fitting}). Using PLD, the planet signal is robustly detected. Our best fit planet model is shown in Figure \ref{fig:k2-43} with 1$\sigma$ uncertainties. The full planet parameters for our best fit joint model of \textit{K2-43b} and \textit{c} are given in Table \ref{tab:planetparams}. We find \textit{K2-43c} to have a radius of 2.42 $R_{earth}$, and an orbital period of 2.42 days, resulting in an equillibrium temperature of $1000K$.

Using the planet vetting tool \texttt{vespa} \citep[see][]{vespa, vespacode} we are able to assign a False Positive Probability (FPP) to both planets in the \textit{K2-43} system. Similarly to many other works (including \cite{dressing} and \cite{crossfield}), we adopt a FPP of $<$1\% to validate a planet. We use direct imaging contrast curves from the NIRC2 instrument at Keck II (obtained through ExoFOP) to rule out neighbouring blended targets. Using the light curves we have generated (see Figure \ref{fig:k2-43}), and stellar parameters derived by \cite{dressing} for \textit{K2-43}, we find an FPP of $\lt 1e{-6}$ for \textit{K2-43b}, and $0.00165$ for \textit{K2-43c}. Additionally, as discussed in \cite{multiboost} and \cite{multik2}, we can apply a ``multiplicity boost'' to our probabilities, as systems with multiple planets are more likely to be true planets. Assuming the same multiplicity boost inferred for the \kepler mission \citep{multiboost}, the FFP for \textit{K2-43c} drops to $6.6104e{-5}$. Based on the small FPP, the absence of any known background stars, and a confirmed planet in the \textit{K2-43} system, we label \textit{K2-43c} a confirmed planet. We also used the Discovery and Vetting of Exoplanets tool (DAVE) \citep{DAVE} to ensure there were no further signatures in the light curve that might indicate a False Positive. We find no significant secondary eclipses for \textit{K2-43c} and no significant odd-even differences between transits. As such we confirm \textit{K2-43c} as a true planet.


\cite{dressing} highlighted that \textit{K2-43b} is a good candidate for atmospheric observations with HST and, in future, JWST, due to the cool host star ($\sim$ 3800K) which is relatively bright in K-band (K $\sim$ 11.5). \textit{K2-43b} and \textit{K2-43c} are on short periods, making observations more easily schedulable. The \textit{K2-43} system could be an excellent candidate for comparative exoplanet atmosphere studies in a single system. However, as with many cool stars, the host star is a significantly spotted star, with flux variations of up to 5\%, which may contaminate atmospheric transmission spectrum.

\subsubsection{What can we learn for TESS?}
Similarly to \ktwo, the \tess spacecraft is also subject to some amounts of motion due to spacecraft jitter (see the Tess Instrument Handbook \citep{jitter}). Unlike the \ktwo roll motion, this jitter causes targets to move randomly on the focal plane, with magnitude of $\lesssim 1$ pixel. In this case, the SFF detrending method will fail, as the motion on the detector is random, and a prediction for the flux at a given pixel location can not be estimated. The PLD method is not dependent on estimating the centroid position of the PSF, or fitting an arclength to its motion. Instead, PLD relies solely on the correlation between changing flux intensity measured by pixels on the detector. This allows PLD to extract correlated noise signals caused by random motion with sub-pixel magnitude, and has been shown to effectively remove jitter noise from \spitzer observations \citep{deming}. In order to improve light curve quality for planet hunting with \tess, we recommend using PLD to detrend spacecraft jitter.

\begin{figure}
    \centering
    \includegraphics[width=1.\textwidth]{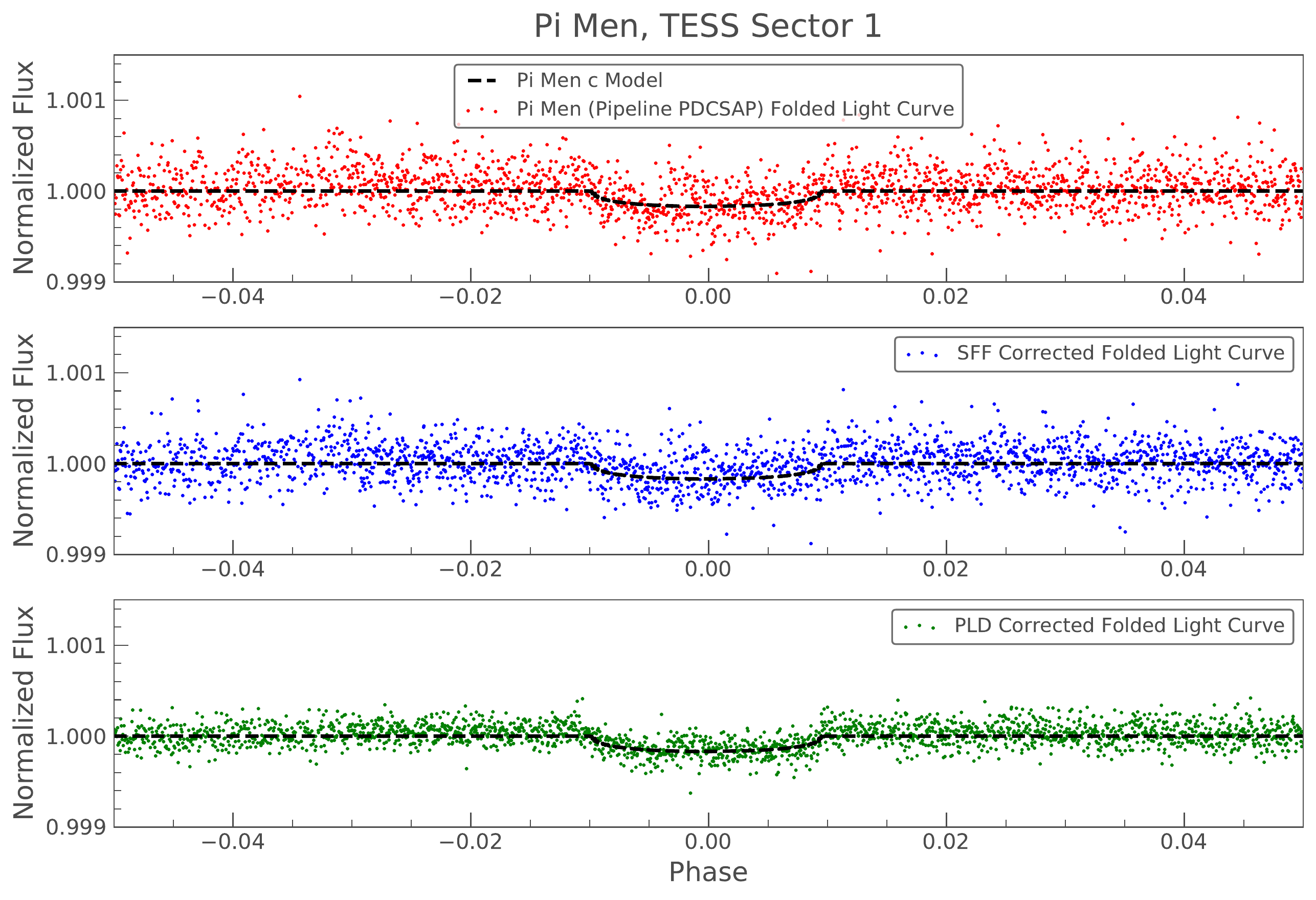}
    \caption{Example of TESS light curve for planet \textbf{pi Men c} from Sector 1. Top: Pre-searched Data Conditioning light curve from the NASA pipeline. Middle: The PDCSAP light curve, corrected using SFF. Bottom: The PDCSAP light curve, corrected using PLD. The PDCSAP and SFF light curves have a comparable scatter, with a Combined Differential Photometric Precision (CDPP) \protect{\citep[see][]{jenkins2010}} of 54 and 58 respectively. The PLD method has better removed the TESS spacecraft jitter, and achieves a CDPP of 32.}
    \label{fig:tessimprovement}
\end{figure}

Figure \ref{fig:tessimprovement} shows an example of a \tess light curve, when \tess jitter is removed. We use confirmed \tess planet \textbf{pi Men c} as an example. The NASA pipeline Pre-search Data Conditioning light curve, an SFF correction and a PLD correction are shown. The reduction in noise for TESS is $>$50\% if PLD is used, compared to using either the NASA pipeline-provided light curve or an SFF detrending. This shows the benefit of removing \tess jitter, in particular by employing the PLD method.

Open source implementations of systematic removal methods are now readily available. The \lk package provides an open source, Python interface for working with \kepler, \ktwo, and \tess data. This includes data access from the MAST archive, the generation of light curves from the calibrated image data, and removal of common instrument systematics. \lk provides a way for users to implement simple versions of both SFF and PLD on any of these data sets, and remove systematics such as the K2 roll motion and \tess jitter. \lk is also tuneable, allowing users to iterate over detrending parameters to establish the effect of the detrending on their final light curve.

\subsection{K2-168}
\label{sec:k2-168}

\begin{sidewaysfigure}
    \centering
    \includegraphics[height=0.45\textwidth]{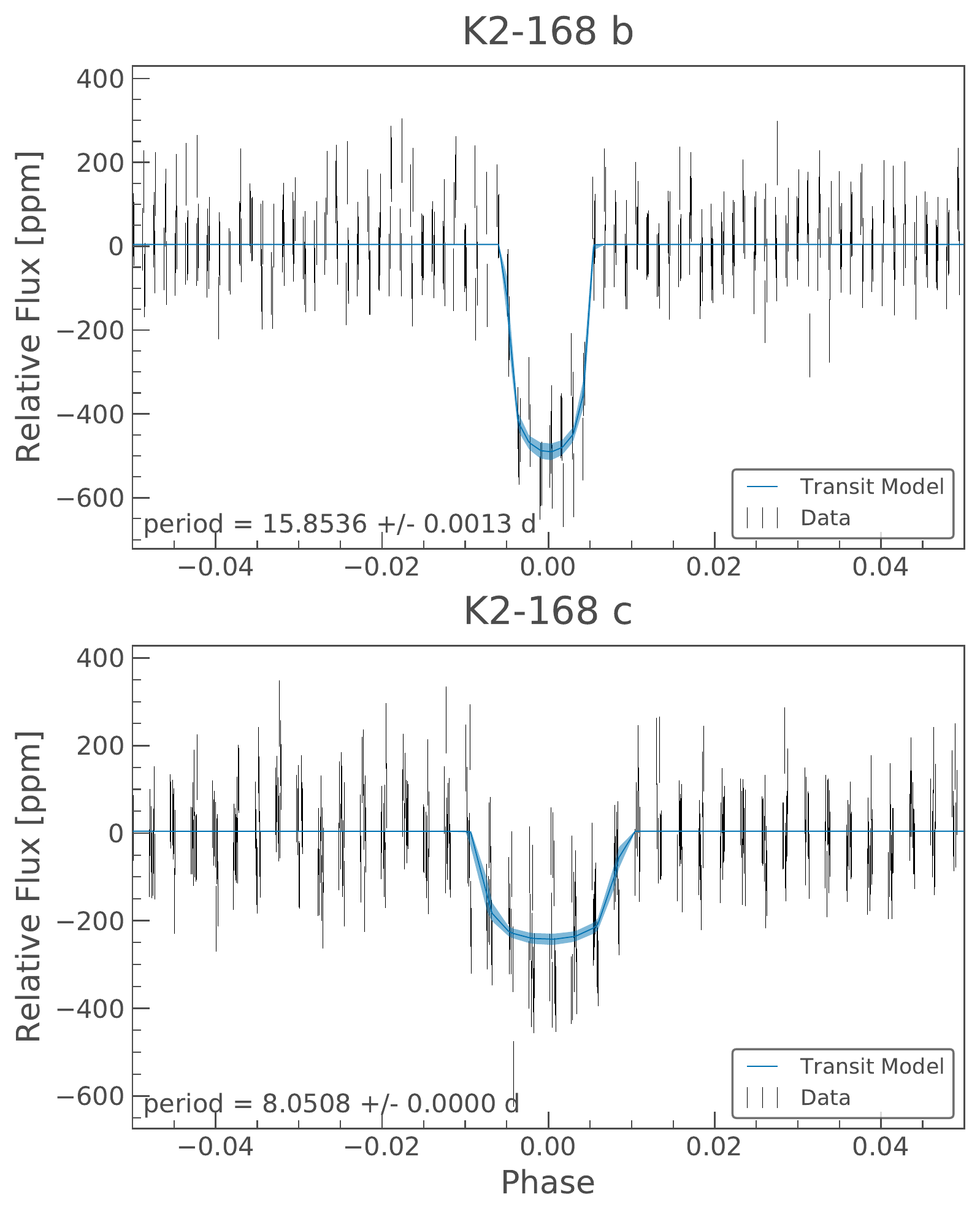}
    \includegraphics[height=0.45\textwidth]{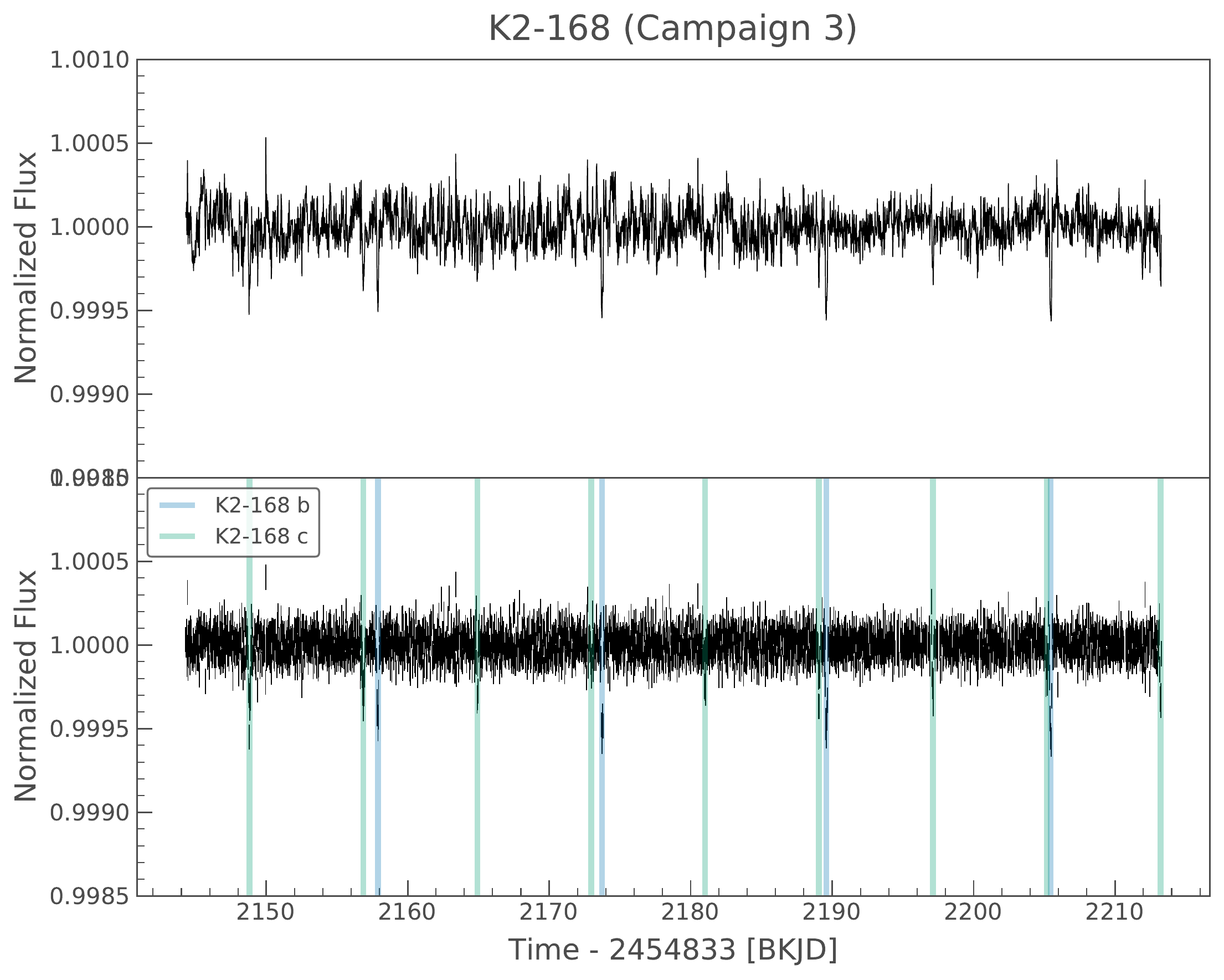}
    \caption{\textit{Left}: Folded transits of \textit{K2-168b} and \textit{c}, having simultaenously removed \ktwo motion systematics and long term stellar variability (see Section \ref{sec:fitting}). Our best fit planet model is show in blue with the 1$\sigma$ uncertainties. \textit{Right}: Light curves of \textit{K2-168}. \textit{Top}: Light curve with motion systematics corrected. \textit{Bottom}: Light curve with both motion systematics and stellar variability removed. Transits of \textit{K2-168b} and \textit{c} have been highlighted. Most of the transits of \textit{K2-168b} occur close to  transit of \textit{K2-168c}.}
    \label{fig:k2-168}
\end{sidewaysfigure}

\textit{K2-168} (EPIC 205950854) was observed in long cadence in \ktwo Campaign 3 (program PI's: Petigura, GO3104; Sanchis-Ojeda, GO3054), and was identified by \cite{Vanderburg2016} and \cite{mayo} to host a super-Earth size planet on a 15.8~day orbit. We have identified a second planet in the system, in a near orbital resonance, with an 8.1~day period.\footnote{After submission of this paper, \cite{heller} identified this signal as a transiting planet candidate using the \tls algorithm \cite{hippke}. The TLS algorithm fits a true transit shape rather than the BLS box, which is similar to the original pipeline process, and can increase the signal to noise of small planet candidates \citep[for discussion see the Kepler Data Processing Handbook;][]{kdph}.} Figure \ref{fig:k2-168} shows our best fit of both planets. The full planet parameters for our best fit joint model of \textit{K2-168b} \textit{c} are shown given in Table \ref{tab:planetparams}. The transit signal of this new planet is detectable in all three of the publicly-available community light curve databases that are hosted at MAST (K2SFF, EVEREST, and K2SC). However, the transit of \textit{K2-168c} can be low signal to noise in some planet discovery pipelines because of its near resonance with \textit{K2-168b}.

The period of \textit{K2-168c} is almost exactly half the period of \textit{K2-168b}, ($P_b/P_c=1.969$), causing the BLS peak for \textit{K2-168c} to occur close to a harmonic peak from \textit{K2-168b}. If, when searching for multiple transiting planets, these resonant peaks were removed, this planet would go undetected. Furthermore, the transits of \textit{K2-168c} also occur close to transits of \textit{K2-168b} in time. If transits of \textit{K2-168b} were removed before searching for multiple transiting planets, many of the transits of \textit{K2-168c} would also be removed. This rare situation, where a second resonant planet is also close in phase, would be alleviated in mission with a longer baseline, where the planets would eventually drift apart. However, for a short baseline mission such as \ktwo, it is possible to have most of the transits overlap, causing shallow transits of resonant planets to remain undetected.

In some planet hunting pipelines, (including \cite{Vanderburg2016}, \cite{mayo}, and the original \kepler pipeline \citep{jenkins2010}) transits are clipped with a small margin before performing a second transit search, in order to identify multi-planet systems. For example, both \cite{Vanderburg2016} and the original \kepler Pipeline remove points within 1.5 days of the transit mid point. This practice can reduce the signal to noise of planets such as \textit{K2-168c}, where transit clipping K2-168b would remove the number of observed transits from 9 to 6. Following the method in \cite{Vanderburg2016}, the SNR of the K2-168c transit is 5.8 if transits are clipped, and 7.2 if transits are not clipped. \cite{Vanderburg2016} and \cite{mayo} require an SNR of 9 before a signal is considered a planet candidate, and so this small signal would not have been detected in either pipeline, even if transits were not clipped. However, this modest increase in SNR and number of transits observed can easily be the difference between a planet signal being triggered in a pipeline, and it remaining undetected.

Using the same method as for \textit{K2-43}, we have calculated the FPP for \textit{K2-168b} and \textit{c}. No contrast curves are available for \textit{K2-168}, and so our FPP values are slightly higher. Using the light curves we have generated (see Figure \ref{fig:k2-168}), and the stellar parameters derived by \cite{mayo} for \textit{K2-168}, we find an FPP of $4.719e{-5}$ for \textit{K2-168b}, and $0.0204$ for \textit{K2-168c} using \texttt{vespa}. While the FPP for \textit{K2-168c} is slightly larger than 1\%, when we apply the multiplicity boost from \cite{multiboost} we find a FPP of $0.00083$ for \textit{K2-168c}. Owing to the multiplicity boost, we can label \textit{K2-168c} a confirmed planet. We also ran vetting tools from the DAVE pipeline and find no significant secondaries or odd-even transit depth differences for \textit{K2-168}. We also find no significant centroid shifts, which would indicate the signal originates from a background star.

Resonant multi-planet systems are particularly valuable for exoplanet science. These systems are likely to exhibit measurable transit timing variations (TTVs), which enable the derivation of mass estimates for planets. \textit{K2-168} is a system close to a 2:1 resonance, suggesting it may exhibit TTVs, though we do not detect significant evidence of TTVs during the short $\sim$ 80-day \ktwo observation. Additionally, both planets are close to the radius gap observed in \kepler, where literature has noted a deficit of planets with $R \sim 1.5 - 2 R_{earth}$ \citep{fulton}. \textit{K2-168} may provide an interesting insight into planet formation close to this boundary. It is important that resonant systems are identified, particularly in missions such as \ktwo and \tess, where planet formation around a variety of stellar types can be understood.

\subsubsection{What can we learn for TESS?}
Like \ktwo, the \tess mission has a short baseline of observation for the majority of targets. While those targets in the continuous viewing zone will have a baseline of up to 1 year, most targets in the \tess prime mission have continuous observations for only 27 days. Systems similar to \textit{K2-168}, where transits from resonant small planets are buried in the harmonics of larger planets, will be difficult to separate in \tess. Extra care should be taken to identify these valuable systems. 75\% of \tess targets will be observed for only a single 27 day sector, and 95\% of \tess targets will be observed for three sectors or less \citep[see][]{barclay}. As such, the majority of \tess targets will have a short baseline, and resonant multi-planet \tess systems will be susceptible to this problem.

The new \exoplanet Python package enables users to simultaneously fit multiple planet systems, benefiting from the fast, stable, and differentiable analytical transit models provided by the \starry package. \exoplanet is able to simultaneously fit transits, long term instrument systematics, and long term stellar variability using Gaussian processes. The loss of additional resonant planets can be avoided by employing \exoplanet to robustly fit and remove exoplanet transits before searching for signatures of additional planets, rather than removing harmonics from the BLS power spectrum or removing transits from the light curve.

\subsection{K2-198}
\label{sec:k2-198}

\begin{sidewaysfigure}
    \centering
    \includegraphics[height=0.45\textwidth]{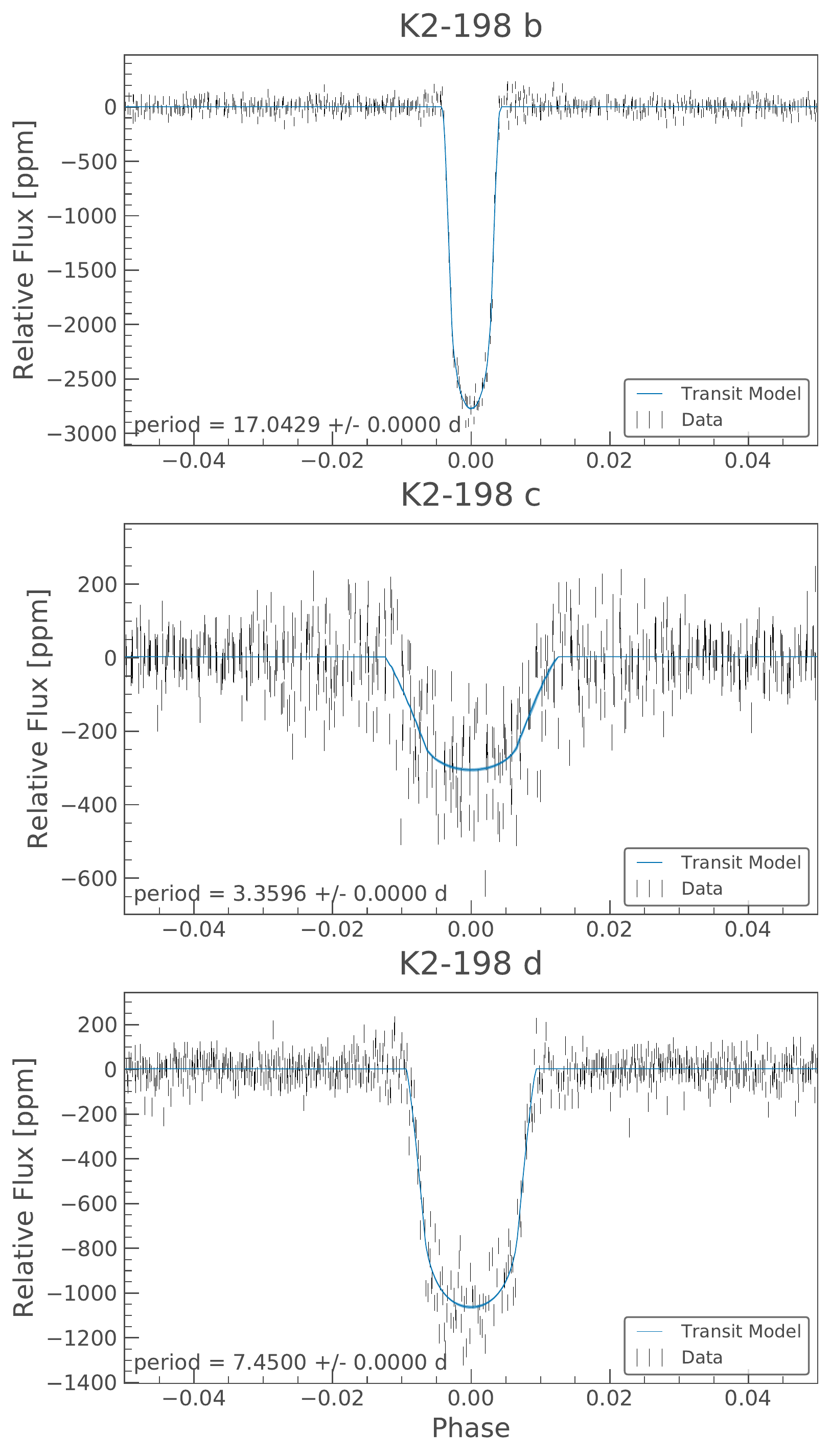}
    \includegraphics[height=0.45\textwidth]{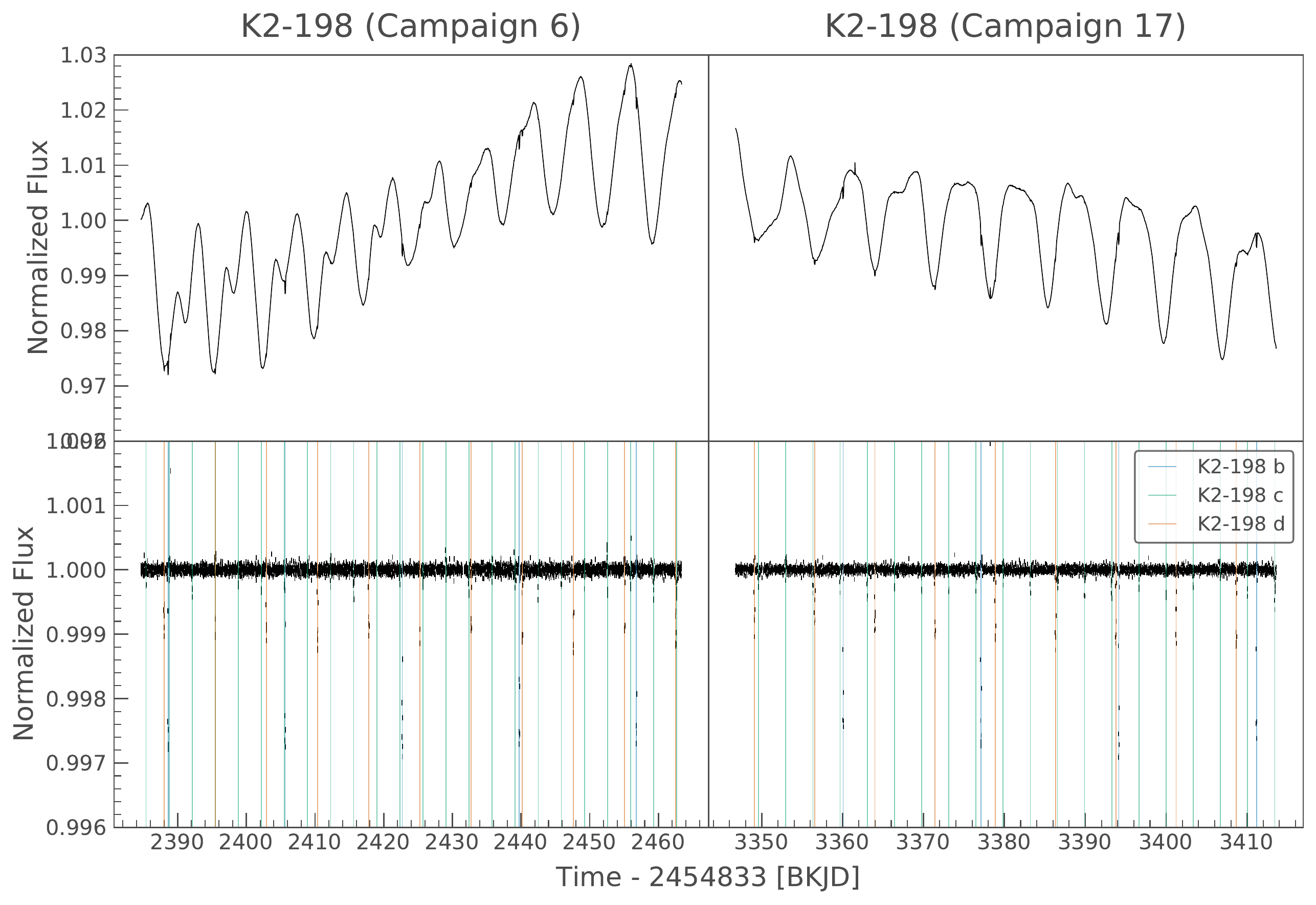}
    \caption{\textit{Left}: Folded transits of \textit{K2-198b, c} and \textit{d}, having simultaenously removed \ktwo motion systematics and long term stellar variability (see Section \ref{sec:fitting}. Our best fit planet model is show in blue alongside the 1$\sigma$ uncertainty. \textit{Right}: Light curves of \textit{K2-198}. \textit{Top}: Light curve with motion systematics corrected. Strong stellar varaibility due to spots is clearly evident. \textit{Bottom}: Light curve with both motion systematics and stellar variability removed. Transits of \textit{K2-198b, c} and \textit{d} have been highlighted.}
    \label{fig:k2-198}
\end{sidewaysfigure}

\textit{K2-198} (EPIC 212768333) was observed in long cadence in \ktwo Campaign 6 (program PI's: Jackson, GO6029; Howard, GO6030; Stello, GO6032; Charbonneau, GO6069; Thompson, GO6086; Dragomir, GO6087).
\cite{mayo} identified a large $\sim$ 4\,$R_{earth}$ exoplanet. The system was also observed later, in Campaign 17. \textit{K2-198} exhibits significant stellar variability, likely from star spots (see Fig. \ref{fig:k2-198}). This spot modulation is approximately $3\%$ in amplitude, much larger than the signal of the transits. If this modulation is not adequately removed, it will suppress signals from a transiting planet.

A common approach to removing stellar variability is to use some form of smoothing filter (e.g., a median or Savitzky-Golay filter). For example, \cite{mayo} and \cite{Vanderburg2016} use a basis spline to remove long term trends. These approaches are fast and simple, making it ideal for planet searches where thousands of light curves must be whitened to remove stellar variability before a planet finding algorithm (e.g., BLS) can be applied. However, such smoothing kernels have two key drawbacks. First, these kernels smooth over any existing planet transits that have not been masked, reducing their transit depth. Second, in cases where planet transits are masked, these kernels can not be used to predict the stellar variability during masked times. As such, there is no way to mask transits and accurately preserve their transit depths. Using Gaussian Processes to accurately model stellar variability overcomes these two issues (see \cite{celerite} for a discussion of the application of GPs in the time domain). GPs are able to robustly predict the stellar variability during masked cadences. Furthermore, it is possible to marginalize over the uncertainties in the best fit GP model hyperparameters, enabling users to accurately estimate the uncertainties due to model uncertainty during transits. As such, a Gaussian Process is a more reliable method for removing stellar variability, prior to searching for transits. While overcoming stellar variability is a common problem, attempted by all planet hunting pipelines, the use of Gaussian Processes to remove this varaibility is quite new, and new tools are now available to the community to easily implement this technique (e.g. \celerite).

In the case of \textit{K2-198}, we have identified two small planets at shorter periods than \textit{K2-198b}, that are only detectable when the long term stellar variability is removed using a GP. Planets \textit{c} and \textit{d} were not identified in \cite{mayo}, who employed a basis spline correction at 1.5 days to remove long term trends. While not reported on the NASA Exoplanet Archive, \cite{pope} identified both \textit{c} and \textit{d} as candidate transits, by employing Gaussian Process detrending to remove stellar variability.

Figure \ref{fig:k2-198} shows the folded transits of all planets in the \textit{K2-198} system, and the removed best-fit stellar variability model, using a Gaussian Process with a Mat\'ern 3/2 kernel. We fit our long term stellar trend simultaneously with our motion systematics removal (see Section ~\ref{sec:fitting}) and robustly propagate the uncertainties by marginalizing over our GP hyperparameters. Figure~\ref{fig:k2-198} clearly shows the transits are detectable only after the stellar variability is accurately removed. The full planet parameters for our best fit joint model of \textit{K2-198b}, \textit{c} and \textit{d} are shown given in Table \ref{tab:planetparams}. \emph{K2-198} was observed in both K2 Campaign 6 and Campaign 17. After we identified these three transits in the Campaign 6 data, we added the Campaign 17 data in our improved light curve (see Section \ref{sec:fitting}) in order to obtain the best planet parameter fits. We find \textit{c} and \textit{d} have radii of 1.4 and 2.4 $R_{earth}$ respectively, with orbital periods of 3.35 and 7.45 days.

Using the same method as for \textit{K2-43} above, we have calculated the FPP for the \textit{K2-198} system. We use contrast curves from direct imaging to rule out close background stars, and inform our FPP prediction. Contrast curves for \textit{K2-198} from the PHARO instrument at the Palomar telescope \citep{pharo} and the DSSI instrument at the WIYN telescope \citep{DSSI} are available on ExoFOP. Using the light curves we have generated (see Figure \ref{fig:k2-198}), and stellar parameters derived by \cite{mayo} for \textit{K2-198}, we find an FPP of $\lt 1e{-6}$ for \textit{K2-198b}, $0.000425$ for \textit{K2-198c}, and $\lt 1e{-6}$ for \textit{K2-198d} using \texttt{vespa}. Without employing any multiplicity boost, we are able to label $K2-198c$ and $d$ confirmed planets. We used the DAVE pipeline to find that there are no significant secondaries or odd-even transit depth differences for \textit{K2-198d}. While the pipeline failed to process \textit{K2-198c}, we consider if to be a confirmed planet based on its low FPP.

\textit{K2-198} is at least a three planet system. Multiple systems with large numbers of planets are particularly useful to test planet formation models and understand the dynamics of planetary systems. Additionally, similarly to \textit{K2-168c}, \textit{K2-198c} is close to the exoplanet radius boundary, increasing the system's potential to test planet formation models, making \emph{K2-198} a valuable system.

\subsubsection{What can we learn for TESS?}
Stellar variability is likely to be a confounding factor for many important \tess planet discoveries. In particular, valuable planet discoveries around young stars are likely to be obscured by large star spots (e.g., see the recent \ktwo discovery of a planet around a variable young star \cite{david}). Planets around small, cool stars, (which are excellent targets for atmospheric characterization with JWST), are also likely to have large stellar variability. Removing this stellar variability accurately is crucial to planet searches around these stars. The original \kepler mission targeted primarily solar-like stars, which are naturally less variable, and vary on longer time scales, making stellar variability less problematic than for the \ktwo and \tess missions. The \ktwo and \tess catalogs contain many more young and active stars, where Gaussian Processes can be highly beneficial for removing stellar variability.

There are several open source packages to mitigate stellar variability that are relevant for \kepler, \ktwo, and \tess. In particular, the \celerite package provides a fast implementation of a Gaussian Process for 1D time series data. \celerite can be used as an alternative to simple smoothing filters to remove long term stellar variability and improve light curves for exoplanet hunting. Additionally, the \exoplanet package provides utilities to fit GP models simultaneously with transit models, allowing the uncertainties associated with fitting the stellar variability to be propagated into the planet parameters.


\section{Conclusions}
\label{sec:conclusions}

In this letter we discuss and confirm 4 \ktwo planets, and provided vetting statistics to confirm them. Each planet was found in a \ktwo system which was already known to host a confirmed planet. We have used \texttt{vespa} to confirm these planets and find low False Positive Probabilities for each planet. Analysis with DAVE shows that there are no significant secondary or tertiary eclipses, odd-even differences between consecutive transits, or photocenter shifts during transits for \textit{K2-43 c}, \textit{K2-168 c} and \textit{K2-198 d}, confirming the planetary interpretation of the transits. Our discoveries suggest that there are still many planets waiting to be found in the K2 data set. Two of the planets we have presented here are close to the planet radius gap \citep{fulton}, thus increasing the value of the \ktwo planet sample to study this gap.

In this work we have performed a search for low SNR transiting planet signals in known K2 planet systems. From this search, we identified four candidates that were not reported in the NASA Exoplanet Archive, and showcased key problems that have confounded exoplanet searches in \ktwo. In this work we have confirmed each planet, discussed the confounding factors in detail, and signposted new open-source tools that can be used to overcome each problem. Our search was not designed to be complete, and was instead designed to highlight new methods that can be easily implemented to find new planets in the \ktwo dataset, and by extension the \tess dataset.

We find that the following factors confounded transit searches for the three systems presented here: 1) Instrument motion systematics (such as the K2 roll motion), particularly those on short time scales of $\lesssim 1$ cadence, can cause noise that buries planet signals. These systematics can be carefully removed in conjunction with stellar systematics in order to reach the highest possible precision (see section \ref{sec:k2-43} and \ref{sec:fitting}), 2) multi-planet systems where planets naturally occur close to resonances can be difficult to identify, particularly if the baseline of the observation is comparable to the orbital period of the planets (see Section \ref{sec:k2-168}) and, 3) The long term stellar variability of the host star can obscure the planet transit signals, if not removed using an appropriate method (see section ~\ref{sec:k2-198}). Each of these key factors that have affected the \ktwo systems presented here are equally important for the recently commissioned \tess mission.

The data used to find these planets has been analyzed by several teams, and processed by several planet hunting pipelines. However, these signals have evaded detection by major pipelines. It has also been shown by previous work using the \kepler detection efficiency that there are more planets to be found in the \ktwo data set \citep[e.g.,][]{dotson}, and that multi-planet systems have been under-reported in the \kepler data set \citep[see][]{christiansen}. The discovery of these planets in known systems highlights that there may indeed be many more planets waiting to be found in the \ktwo data, if the challenges discussed in this work are addressed. While \kepler focused primarily on quiet stars, where stellar variability is a lower magnitude effect, gains could potentially still be made by robustly removing stellar variability with a GP before planet searching in the \kepler dataset.

These challenges can be addressed using new open source community tools. Two common motion systematics mitigation techniques (SFF and PLD) are now available in the new \lk toolkit. The new fast implementation of BLS now in available in \bls, can be used in conjunction with \lk to iteratively search for planets while varying systematics removal parameters, and masking known transits. The new \exoplanet package, based on the \starry and \celerite, can implement fast exoplanet transit modeling and the removal of stellar variability using GPs.  These toolkits are ready to be applied to both \ktwo and \tess archival data to find empower a comprehensive survey of high-value, small planets in multi-planet systems, further expanding the legacy of NASA's exoplanet finding missions.

\section{Acknowledgements}
This paper uses several community tools, and we would like to thank all of the authors of these community packages for documenting and releasing their code. In particular, we would like to recognize the exceptional efforts by Daniel Foreman-Mackey, Rodrigo Luger, and Timothy Morton. This paper includes data collected by the K2 mission and obtained from the MAST data archive at the Space Telescope Science Institute (STScI). Funding for the K2 mission is provided by the NASA Science Mission Directorate. STScI is operated by the Association of Universities for Research in Astronomy, Inc., under NASA contract NAS 5–26555. This research has made use of the NASA Exoplanet Archive, which is operated by the California Institute of Technology, under contract with the National Aeronautics and Space Administration under the Exoplanet Exploration Program. This paper has made use of the ExoFOP service, which is funded by NASA through the NASA Exoplanet Science Institute.

\facility{Kepler, MAST, Exoplanet Archive}

\begin{table}[b]
\centering
\begin{tabular}{lll}
\toprule
{} &                             \emph{K2-43 b} &                          \emph{K2-43 c} \\
\midrule
Period [days]                &     3.471149 $\pm _{0.000052} ^{0.000104}$ &  2.198884 $\pm _{0.000011} ^{0.000022}$ \\
Transit Midpoint [JD]        &  2456809.8843$\pm_{-0.00071}^{0.00071}$ &     2456810.4059$\pm_{-0.0098}^{0.0099}$ \\
Radius [$R_{earth}$]         &                 4.51 $\pm _{0.19} ^{0.44}$ &              2.42 $\pm _{0.11} ^{0.26}$ \\
R$_P$/R$_*$                  &               0.04298$\pm_{-0.001677}^{0.002318}$ & 0.02319$\pm_{-0.001004}^{0.001296}$ \\
Impact Parameter             &                 0.11 $\pm _{0.17} ^{0.32}$ &              0.14 $\pm _{0.18} ^{0.32}$ \\
Inclination [degrees]        &              89.60 $\pm _{-0.67} ^{-1.35}$ &           89.26 $\pm _{-0.95} ^{-1.85}$ \\
Semi Major Axis [a/R$^*]$    &                 8.00 $\pm _{0.25} ^{0.49}$ &              5.90 $\pm _{0.18} ^{0.36}$ \\
Equillibrium Temperature [K] &                939.3 $\pm _{18.5} ^{37.7}$ &            1093.7 $\pm _{21.5} ^{43.9}$ \\


\midrule
Stellar Mass [Msol]               &  0.571 $\pm _{0.055} ^{0.111}$ \citep{dressing}&\\
Stellar Radius [Rsol]             &  0.542 $\pm _{0.022} ^{0.049}$ \citep{dressing}&\\
Stellar Effective Temperature [K] &   3840.6 $\pm _{49.5} ^{98.8}$ \citep{dressing}&\\
Limb Darkening 1          &  0.031 $\pm _{0.073} ^{0.185}$ &\\
Limb Darkening 2          &  0.054 $\pm _{0.141} ^{0.390}$ &\\
\bottomrule
\end{tabular}
\vspace{20pt}

\begin{tabular}{lll}
\toprule
{} &                         \emph{K2-168 b} &                         \emph{K2-168 c} \\
\midrule
Period [days]                &       15.8523 $\pm _{0.0012} ^{0.0025}$ &  8.050722 $\pm _{0.000038} ^{0.000075}$ \\
Transit Midpoint [JD]        &  2456975.03774$\pm_{-0.0023}^{0.0023}$ &     2456973.7648$\pm_{-0.0099}^{0.0100}$ \\
Radius [$R_{earth}$]         &              1.86 $\pm _{0.11} ^{0.24}$ &           1.310 $\pm _{0.063} ^{0.144}$ \\
R$_P$/R$_*$                  &               0.01812$\pm_{-0.0010}^{0.001}$ & 0.01258$\pm_{-0.00057}^{0.00075}$ \\
Impact Parameter             &              0.32 $\pm _{0.11} ^{0.20}$ &           0.064 $\pm _{0.141} ^{0.307}$ \\
Inclination [degrees]        &           89.40 $\pm _{-0.25} ^{-0.46}$ &           89.81 $\pm _{-0.43} ^{-0.98}$ \\
Semi Major Axis [a/R$^*]$    &             25.42 $\pm _{0.28} ^{0.57}$ &             16.18 $\pm _{0.18} ^{0.36}$ \\
Equillibrium Temperature [K] &           766.35 $\pm _{8.12} ^{16.33}$ &             960.6 $\pm _{10.2} ^{20.5}$ \\
\midrule

Stellar Mass [Msol]               &  0.877 $\pm _{0.029} ^{0.060}$ \citep{mayo}&\\
Stellar Radius [Rsol]             &  0.830 $\pm _{0.043} ^{0.087}$ \citep{mayo}&\\
Stellar Effective Temperature [K] &  5502.7 $\pm _{50.5} ^{101.0}$ \citep{mayo}&\\
Limb Darkening 1          &     0.10 $\pm _{0.19} ^{0.40}$ &\\
Limb Darkening 2          &  0.083 $\pm _{0.206} ^{0.501}$ &\\
\bottomrule

\end{tabular}

\vspace{20pt}

\begin{tabular}{llll}
\toprule
{} &                             \emph{K2-198 b} &                            \emph{K2-198 c} &                            \emph{K2-198 d} \\
\midrule
Period [days]                &  17.0428683 $\pm _{0.0000035} ^{0.0000071}$ &  3.3596055 $\pm _{0.0000021} ^{0.0000040}$ &  7.4500177 $\pm _{0.0000026} ^{0.0000052}$ \\
Transit Midpoint [JD]        &   2457204.5687$\pm_{-0.00014}^{0.00014}$ &     2457215.0320$\pm_{-0.0017}^{0.0019}$ &    2457213.5759$\pm_{-0.0010}^{0.0010}$ \\
Radius [$R_{earth}$]         &               4.189 $\pm _{0.098} ^{0.228}$ &              1.423 $\pm _{0.036} ^{0.081}$ &              2.438 $\pm _{0.056} ^{0.130}$ \\
R$_P$/R$_*$                  &               0.039$\pm_{-0.00088}^{0.0010}$ & 0.013$\pm_{-0.00032}^{0.00037}$   & 0.022$\pm_{-0.00050}^{0.00057}$ \\
Impact Parameter             &            0.6610 $\pm _{0.0099} ^{0.0320}$ &              0.715 $\pm _{0.012} ^{0.030}$ &              0.047 $\pm _{0.104} ^{0.226}$ \\
Inclination [degrees]        &            88.904 $\pm _{-0.027} ^{-0.094}$ &           86.494 $\pm _{-0.088} ^{-0.268}$ &              89.86 $\pm _{-0.30} ^{-0.68}$ \\
Semi Major Axis [a/R$^*]$    &                 25.86 $\pm _{0.48} ^{0.95}$ &                 8.76 $\pm _{0.16} ^{0.32}$ &                14.90 $\pm _{0.28} ^{0.55}$ \\
Equillibrium Temperature [K] &               715.80 $\pm _{9.26} ^{18.72}$ &               1229.9 $\pm _{15.9} ^{32.2}$ &                943.2 $\pm _{12.2} ^{24.7}$ \\
\midrule
Mass [Msol]               &  0.799 $\pm _{0.045} ^{0.091}$ \citep{mayo}&&\\
Radius [Rsol]             &  0.757 $\pm _{0.016} ^{0.035}$ \citep{mayo}&&\\
Effective Temperature [K] &   5212.9 $\pm _{49.2} ^{99.0}$ \citep{mayo}&&\\
Limb Darkening 1          &  0.303 $\pm _{0.073} ^{0.146}$ &&\\
Limb Darkening 2          &     0.25 $\pm _{0.11} ^{0.23}$ &&\\

\bottomrule
\end{tabular}
\vspace{20pt}

\caption{Best fit parameters for all known planets in the \textit{K2-43}, \textit{K2-168}, and \textit{K2-198} systems. See Section \ref{sec:fitting} for details on the planet fitting procedure. We find consistent parameters with literature results for previously confirmed planets \textit{K2-43b, K2-168b,} and \textit{K2-198b}.}
\label{tab:planetparams}
\end{table}

\bibliographystyle{aasjournal}
\bibliography{bibliography.bib}{}

\end{document}